\documentclass[twocolumn,aps,pre,superscriptaddress,floatfix,a4paper,showkeys]{revtex4-2}

\usepackage[T1]{fontenc}
\usepackage[utf8]{inputenc}
\usepackage{amsmath}
\usepackage{amssymb,bbold}
\usepackage{mathtools}
\usepackage{siunitx}

\makeatletter
\renewcommand{\p@subsection}{}
\renewcommand{\p@subsubsection}{}
\makeatother

\newcommand{\bfhat}[1]{\mathbf{\hat #1}}

\newcommand{\mbf}[1]{\mathbf #1}

\newcommand*\chem[1]{\ensuremath{\mathrm{#1}}} 

\begin{document}
	
	
	\title{Magnetic elastomers as specific soft actuators -- predicting particular modes of deformation from selected configurations of magnetizable inclusions}
	
	\author{Lukas Fischer}
	\email{lukas.fischer@ovgu.de}
	\affiliation{Institut f{\"u}r Physik, 
		Otto-von-Guericke-Universität Magdeburg, Universitätsplatz 2, 39106 Magdeburg, Germany}
	
	\author{Andreas M. Menzel}
	\email{a.menzel@ovgu.de}
	\affiliation{Institut f{\"u}r Physik, 
		Otto-von-Guericke-Universität Magdeburg, Universitätsplatz 2, 39106 Magdeburg, Germany}
	
	\date{\today}
	
\begin{abstract}
    Amongst the various fascinating types of material behavior featured by magnetic gels and elastomers are magnetostrictive effects. That is, deformations in shape or changes in volume are induced from outside by external magnetic fields. Application of the materials as soft actuators is therefore conceivable. Mostly, straight contraction or extension of the materials along a certain direction is discussed and investigated in this context. Here, we demonstrate that various further, different, higher modes of deformation can be excited. To this end, different spatial arrangements of the magnetizable particles enclosed by the soft elastic matrix, which constitute the materials, need to be controlled and realized. We address various different types of spatial configurations of the particles and evaluate resulting types of deformation using theoretical tools developed for this purpose. Examples are sheet-like arrangements of particles, circular or star-shaped arrangements of chain-like aggregates, or actual three-dimensional star-like particle configurations. We hope to stimulate with our work the development of experimental design and engineering methods so that selected spatial particle arrangements in magnetic gels and elastomers can be put to reality. Overall, we in this way wish to promote the transfer of these promising class of materials to real-world applications.
\end{abstract}


\maketitle

\section{Introduction}
\label{sec:introduction}

We consider soft materials that consist of a reversibly deformable, elastic, continuous matrix containing discrete, magnetizable inclusions. Magnetic gels and elastomers \cite{jolly1996magnetoviscoelastic,filipcsei2007magnetic,odenbach2016microstructure,weeber2018polymer,menzel2015tuned,lopez2016mechanics,schmauch2017chained,weeber2019studying,stolbov2019magnetostriction,menzel2019mesoscopic,schumann2019microscopic} represent prime examples of this class of materials. In this case, the elastic matrix is of polymeric origin, crosslinked to an elastic network and possibly swollen by a solvent for elevated softness. The inclusions can be regarded as solid particles. For possible experimental realizations, the diameters of these particles are typically of submillimeter range, from tens to a few hundreds of micrometers. 

When magnetized by external magnetic fields, the particles interact with each other magnetically. In the absence of external magnetic field gradients, the interparticle interactions are the only forces generated on the inclusions. The particles are enclosed by the surrounding elastic matrix and cannot move through it due to their size. Therefore, the forces generated on the particles are transmitted to the surrounding soft elastic matrix, leading to its deformation \cite{puljiz2016forces, puljiz2017forces, puljiz2019displacement, raikher2008shape, metsch2016numerical}. 

From such effects occurring on the microscopic particle scale, overall macroscopic material properties are affected. Above all, the macroscopic mechanical material behavior is modified, which is frequently referred to as the magnetorheological effect. More precisely, the induced magnetic interactions lead to changes in the static elastic moduli \cite{filipcsei2007magnetic,pessot2014structural,wood2011modeling,ivaneyko2012effects,evans2012highly,han2013field,borin2013tuning,zubarev2019rheological,volkova2017motion} as well as in the dynamic storage and loss moduli \cite{jolly1996magnetoviscoelastic,jolly1996model,stepanov2007effect,bose2009magnetorheological,jarkova2003hydrodynamics,chen2007investigation,chertovich2010new,chiba2013wide,pessot2016dynamic,sorokin2015hysteresis,pessot2018tunable,gila2018rheology,watanabe2018effect}. Since this effect works in a contactless way by external magnetic field, is tunable by the strength of the external field, and often is reversible \cite{borin2024magnetic}, a multitude of possible applications results. Frequently, magnetically tunable vibration absorbers and damping devices are mentioned in this context \cite{carlson2000mr,deng2006development,sun2008study,liao2012development,molchanov2014viscoelastic,becker2020magnetoactive}. 

It has been demonstrated by microcomputed x-ray tomography that the change in overall mechanical behavior is correlated to internal restructuring \cite{schumann2017characterisation}. Specifically, chain-like aggregates of initially well separated magnetized particles can form by deformation of the surrounding elastic matrix \cite{gundermann2014investigation, gundermann2017statistical, pessot2018tunable}. Not only is the mechanical behavior affected by such internal restructuring. The particles usually feature electric and thermal conductivities that are notably different from those of the surrounding elastic matrix. Thus, when the particles reversibly form anisotropic aggregates (such as chains), this causes magnetically tunable changes in transport properties, especially electric and thermal conductivity \cite{Kchit2009,2Kchit2009,Mietta2016,DiazBleis2014,Pech-May2018,jager2022variations}. Only a few studies have addressed this behavior so far. 

We here consider a third type of overall response of the materials to external magnetic fields. Namely, these are macroscopic magnetostrictive effects \cite{gollwitzer2008measuring,filipcsei2010magnetodeformation,stolbov2011modelling,zubarev2013effect,zubarev2013magnetodeformation,maas2016experimental,attaran2017modeling,guan2008magnetostrictive,gong2012full,vazquez2021composite,keip2017computational,fischer2019magnetostriction,fischer2020,fischer2021}. That is, the deformations induced by the magnetized particles in their surroundings due to the resulting magnetic interparticle interactions result in overall macroscopic distortions. So far, this type of behavior has been discussed in the context of soft magnetoelastic actuators \cite{filipcsei2007magnetic,an2003actuating,zimmermann2007deformable,raikher2008shape,fuhrer2009crosslinking,bose2012soft,schmauch2017chained,hines2017soft,li2014state,lum2016shape,maas2016experimental,becker2019magnetic,vazquez2021composite,chavez2020actuators,vazquez2023fabrication} that can serve, for instance, as magnetic valves. 

It is a significant challenge especially for theoretical studies to link the properties on the microscopic particulate scale, especially the particle configurations and resulting magnetically induced interactions, to the macroscopic scale of overall material behavior \cite{weeber2012deformation,menzel2014bridging, weeber2015ferrogels2,pessot2015towards,metsch2016numerical, menzel2021stimuli, roghani2023effect,kalina2023fe}. In this way, the parameters in macroscopic continuum descriptions \cite{jarkova2003hydrodynamics, bohlius2004macroscopic,gebhart2019general} could be substantiated by their microscopic origin. Several previous works resort to statistical approaches to handle the many microscopic degrees of freedom associated with the magnetizable particles \cite{cremer2017density, romeis2016elongated}. These approaches by construction rely on a certain type of averaging and are justified from a coarse-grained macroscopic perspective for large samples. 
However, after all, each sample of an elastic composite system is of finite size. Moreover, the particle positions are fixed in this type of materials, so that each individual sample by itself does not necessarily show an averaged behavior. If we want to describe quantitatively and in detail the deformation of one specific sample, we need to calculate the overall response from the discrete arrangement of all individual inclusions in this specific realization. 

Therefore, our approach is different from averaging procedures. Based on a Green’s function formalism \cite{walpole2002elastic, fischer2019magnetostriction}, we have developed a theoretical description that allows us to evaluate from the discrete arrangement of magnetizable inclusions and the resulting magnetic interactions when exposed to an external magnetic field the overall macroscopic deformation of the whole system. The approach is valid in the regime of linear elasticity, typically corresponding to not more than moderate elastic deformations (of, for instance, about $10\%$ \cite{puljiz2018reversible}), and for sufficiently separated magnetizable inclusions. We have demonstrated that, on this basis, we can determine for various regular lattice-like arrangements of magnetizable inclusions inside the elastic matrix the resulting, magnetically induced, overall deformations for sphere-like systems \cite{fischer2019magnetostriction, fischer2021}. 

In one first study, we have shown that more specific modes of overall deformation can be excited by an appropriately chosen arrangement of the internal particle structure. More precisely, we considered globally twisted arrangements of chain-like structures and helical particle configurations \cite{fischer2020, menzel2021stimuli}. As a result, an overall torsional mode of deformation was induced. Such systems have recently been realized experimentally as well \cite{vazquez2023fabrication}. 

It is our goal in the present work to extend such types of considerations. Instead of the previously addressed regular lattice-like arrangements, we now turn to more specific structures of magnetizable particles. We demonstrate that an extended spectrum of modes of overall deformation with special focus on certain specific modes of deformation can be obtained by realizing these more specific arrangements of particles. Such a perspective highlights the potential of the materials when searching for soft elastic actuators of peculiar types of induced deformation for individual tasks. 

We proceed by briefly summarizing the background of the theoretical approach that we have developed for this purpose in Sec.~\ref{sec:theory}. After that, we introduce several different types of specific arrangements of particles and evaluate their resulting overall macroscopic magnetoelastic response in the several subsections contained in Sec.~\ref{sec:results}. Our conclusions together with a brief perspective are provided in Sec.~\ref{sec:conclusions}.

\section{Theoretical description}
\label{sec:theory}

If one wishes to calculate the actual deformation of a system as viewed from outside, it is mandatory to include the boundaries explicitly. Any sample is of finite size. When we discuss its deformation in the context of actuation, it is usually the displacement of its surfaces that determines the considered response. 

Calculating by analytical theory the elastic deformations in a finite-sized system provides an extreme challenge. We have found a way by turning to spherical systems.
In this case, building on previous work that considered an elastic sphere enclosed by an infinitely extended elastic matrix \cite{walpole2002elastic}, we managed to determine the associated Green’s function for the elastic deformation of a free-standing elastic sphere \cite{fischer2019magnetostriction}. The Green’s function quantifies the elastic displacements resulting in the elastic sphere in response to point-like force centers acting within the sphere on the elastic material. In our case, we consider the magnetizable inclusions as these point-like force centers when they are magnetized in the external magnetic field. To render such an approach quantitatively valid, the particulate inclusions need to be well separated from each other relative to their diameter. 

More precisely, we consider homogeneous, isotropic, linearly elastic matrices. Their deformations are quantified by the Navier--Cauchy equations \cite{cauchy1828exercises}
\begin{equation}
\label{eq:NC}
\mu\Delta \mathbf{u}(\mathbf{r}) + \frac{\mu}{1-2\nu}\nabla\nabla\cdot\mathbf{u}(\mathbf{r}) = -\mathbf{f}_b(\mathbf{r}).
\end{equation}
In this equation, $\mathbf{u}(\mathbf{r})$ corresponds to the elastic displacement field at position $\mathbf{r}$, $\mu$ is the elastic shear modulus, $\nu$ represents the Poisson ratio that quantifies the compressibility of the elastic matrix, and $\mathbf{f}_b(\mathbf{r})$ sets the bulk force density acting on the elastic material.
To determine the resulting displacement field, we set $\mathbf{f}_b(\mathbf{r})=\sum_{i=1}^N \mathbf{F}_i\delta(\mathbf{r}-\mathbf{r}_i)$ in our case, where $N$ provides the overall number of particulate inclusions, $i$ labels the inclusions, while $\mathbf{F}_i$ specifies the magnetic force acting on the $i$th inclusion and $\mathbf{r}_i$ its position. $\delta(\mbf{r})$ represents the Dirac delta function. An analytical expression for the Green's function associated with Eq.~(\ref{eq:NC}) for a spherical geometry is available \cite{walpole2002elastic, fischer2019magnetostriction}, yet complex and lengthy, so we do not reproduce it here. After derivation of this Green's function for the geometry of a free-standing elastic sphere \cite{fischer2019magnetostriction}, we assume the inclusions as well separated from each other and thus to reasonable approximation as point-like.
Then, we obtain the resulting displacement field as
\begin{equation}
\mathbf{u}(\mathbf{r}) = \sum_{i=1}^N \mathbf{\underline{G}}(\mathbf{r},\mathbf{r}_i)\cdot\mathbf{F}_i,
\end{equation}
where $\mathbf{\underline{G}}(\mathbf{r},\mathbf{r}_i)$ denotes the corresponding Green's function, $\mathbf{r}$ the position at which we evaluate the displacement field, and $\mathbf{r}_i$ the position of the $i$th inclusion. 

We evaluate the resulting displacement field $\mathbf{u}(\mathbf{r})$ at $49152$ well distributed positions on the surface of the elastic sphere, relying on the distribution defined by the HEALPix package \cite{HEALPix}. For our purpose, we then consider the components of radial outward, azimuthal, and polar surface displacements $u^{\bot}(\mathbf{r})$, $u^{\varphi}(\mathbf{r})$, $u^{\theta}(\mathbf{r})$, respectively. These components are directly connected to the deformation of the sphere as viewed from outside.
We then expand these components of the surface displacement field into spherical harmonics. In this way, since spherical harmonics form a complete basis set of orthonormal functions on the surface of the sphere, we thus determine the resulting spectrum of normal modes of deformation. The magnitudes of the expansion coefficients are related to the strengths of the particular types of deformation associated with the specific modes. 

Finally, since the inclusions are well separated from each other, we approximate their mutual magnetic interactions as dipolar. Therefore, the magnetic force $\mathbf{F}_i$ acting on the $i$th inclusion is specified as \cite{jackson1962classical}
\begin{equation}
\label{eq:dipoles}
\mbf{F}_{i}=-\sum_{\substack{j=1 \\ j\neq i}}^N \frac{3 \mu_0 m^2 
		\left[ 5 \bfhat{r}_{ij} \left( \bfhat{m} \cdot \bfhat{r}_{ij}\right)^2 - \bfhat{r}_{ij} - 2 \bfhat{m} \left( \bfhat{m} \cdot \bfhat{r}_{ij} \right)  \right]}{4 \pi r_{ij}^4}.
\end{equation}
In this expression, $\mu_0$ is the magnetic vacuum permeability. We consider all magnetizable inclusions to be identical in size and magnetic properties and exposed to strong homogeneous external magnetic fields, so that they are magnetized to saturation. Then, the magnetic moment $\mbf{m}$ of magnitude $m$ and orientation $\bfhat{m}$ is the same for all these identical magnetic inclusions. In our coordinate frame, we set $\bfhat{{z}}\parallel\bfhat{{m}}$. Moreover, we introduced $ \mbf{r}_{ij} = \mbf{r}_{i} - \mbf{r}_{j}$, ${r}_{ij}=|\mbf{r}_{ij}|$, and $\bfhat{{r}}_{ij}=\mbf{r}_{ij}/r_{ij}$.
Since the deformation of the material influences the positioning of the magnetizable inclusions, which in turn affects the magnetic interactions that cause the deformations in the first place, we use an iterative numerical scheme \cite{fischer2019magnetostriction}. It leads us to the final deformed state and thus to the set of magnetic interaction forces that is utilized to calculate the surface displacements.

\section{Specific discrete arrangements of particulate inclusions and overall mechanical response}
\label{sec:results}

In this part, we investigate the induced mechanical response of the whole system as induced by various different types of spatial configurations of discrete particulate magnetizable inclusions. To respect the requirements in Sec.~\ref{sec:theory}, we impose a minimal center-to-center separation distance of $0.12 R$ of the individual particles from each other, where $R$ is the radius of the elastic sphere that in the following sets our unit of length. Simultaneously, a minimal distance of $0.06 R$ is maintained from the surface of the sphere. We assume spherical inclusions of radius $0.02 R$.
To quantify the strength of the magnetic interactions relative to the elastic interactions, we rely on a nondimensional parameter $ 3 \mu_0 m^2/ 4\pi \mu R^6 $. In the following, its magnitude is set to $ 5.4 \times 10^{-8} $, in line with experimental parameter values \cite{fischer2019magnetostriction}. One possible experimental set of parameters that leads to this value is given by an elastic matrix of shear modulus $\mu \approx \SI{1.67}{\kilo\pascal}$ \cite{filipcsei2010magnetodeformation,an2003actuating,gollwitzer2008measuring} and saturation magnetization of $\SI{518}{\kilo\ampere\per\meter}$ for the magnetizable inclusions, as would be realistic for \chem{Fe_3 O_4} \cite{cornell2003iron}. Besides, the inclusion radius is set to $0.02 R$ as above.
Along these lines, we evaluate in the following the overall magnetoelastic response of various spatial arrangements of chain-like aggregates, hexagonally structured layer-like configurations, spherical arrangements, three-dimensional star-shaped configurations, and single- and double-stranded helical arrangements. It becomes obvious that the different spatial arrangement on the particulate microscale is connected to and in parts reflected by the resulting spectrum of modes of overall deformation. Varying the microscopic structure thus generally enables modifying the relative magnitude of dominating macroscopic modes. 

In each of the various cases considered in the following subsections, we address four different values of the Poisson ratio. They quantify the degree of compressibility of the elastic matrix. Specifically, $\nu = 0.5$ describes completely incompressible materials that do not allow any changes in volume under deformation. Next, $\nu = 0.3$ represents materials that are moderately compressible. $ \nu = 0$ describes a strongly compressible material for which elongation along one axis does not lead to contraction along the perpendicular axes. Lastly, $\nu = -0.5$ is associated with materials that even elongate along the perpendicular axes when stretching it along an initially selected axis. Such astonishing behavior is called auxetic.

\subsection{Chain-like structures}
\label{subsec:chains}

The first kind of arrangements of magnetizable inclusions inside the elastic matrix that we consider are various spatial organizations of chain-like structures. For this purpose, we specify different two-dimensional arrangements in the central plane of the elastic sphere, the normal vector to this plane coinciding with the magnetization direction. These configurations in the central plane set the spatial organization of our chain-like structures. Then, we stack magnetizable inclusions along the magnetization direction above and below these inclusions with a spacing of $0.121 R$ to establish the chain-like aggregates filling the sphere.

\subsubsection{Planar star-shaped arrangements}
\label{subsubsec:star}
\begin{figure}
	\includegraphics[width=\linewidth, trim={4.7cm 0.1cm 5.cm 0},clip]{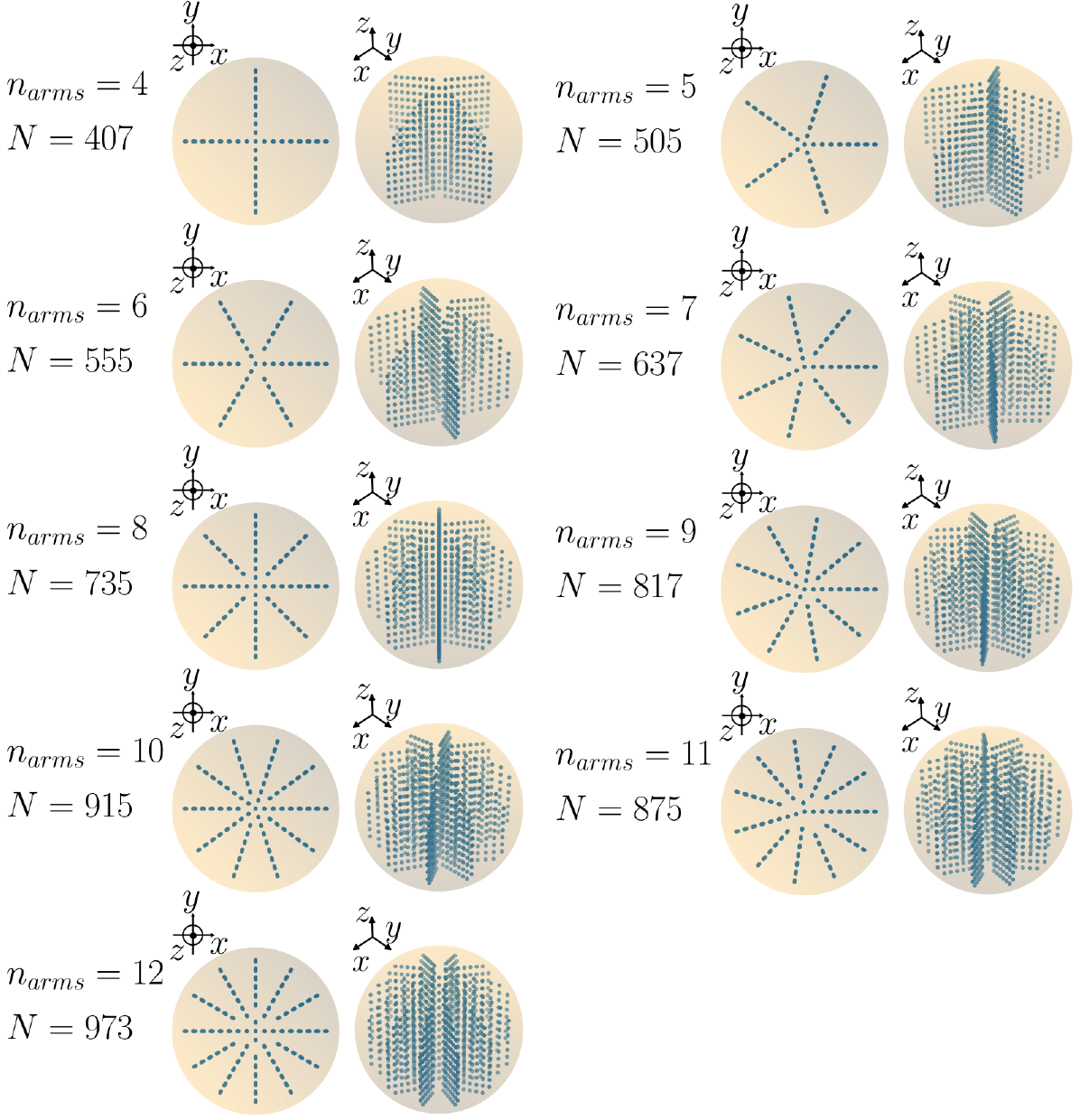}
	\caption{Top views (left of each pair of snapshots) and tilted top views (right of each pair) of the star-shaped arrangements of chains of particles. The chains are oriented along the magnetization direction $\bfhat{z}$. $n_{arms}$ indicates the number of arms of each star and  $N$ sets the number of magnetizable inclusions. Gaps around the centers in the top views in the regular arrangements for $n_{arms} \geq 6$ result from our constraints, see also Appendix A.}
	\label{fig:star_configs}
\end{figure}

In the first case, we use a regular two-dimensional star-shaped arrangement of inclusions in the central plane, where we first select the number of arms $n_{arms}$. The plane is then evenly divided into $n_{arms}$ segments, with inclusions placed on the boundaries between these segments. One inclusion is always placed at the center of the sphere. The chain-like aggregates are then grown towards the top and the bottom of the sphere from the central plane. Along one arm, we use the same distance of nearest neighbors as the distance within the chain-like aggregates ($0.121 R$). Additionally, for neighboring particles in one arm, we use an alternating vertical shift of $0.121 R / 2$. This allows for more inclusions per arm and in total in our final configuration while respecting the imposed minimal distance of neighboring inclusions. For high $n_{arms} \geq 6$, some chains close to the center were deleted because they are too close to each other, in line with our constraints. We visualize top and tilted top views of the resulting arrangements in Fig.~\ref{fig:star_configs}, with the missing chains visible as gaps in the regular arrangements. Additionally, we list the missing chains explicitly in Appendix A.

\begin{figure}
	\includegraphics[width=\linewidth, trim={0.2cm 0.1cm 0.cm 0.cm},clip]{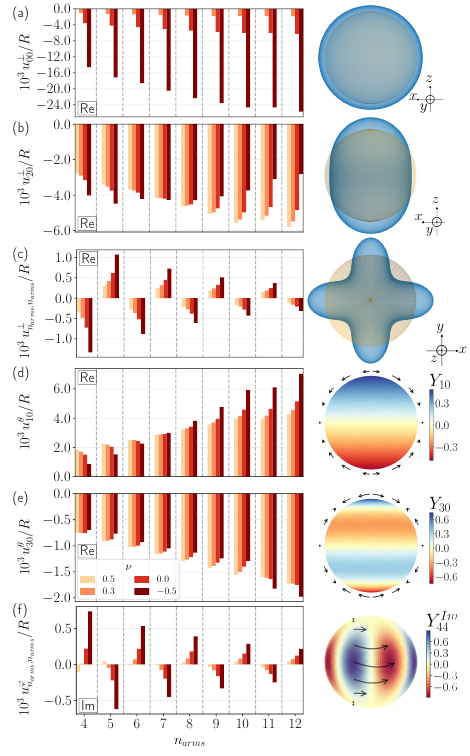}
	\caption{For the star-shaped chain configurations in Fig.~\ref{fig:star_configs}, we evaluate the most important deformational modes. They are characterized by an expansion into spherical harmonics of the three components of the displacement field on the surface of the elastic sphere. On the left-hand side, we plot the largest expansion coefficients depending on $n_{arms}$ and the Poisson ratio $\nu$ of the elastic material. 
    The second index of the respective mode $m$ can correspond to a $\varphi$-dependence of $ \cos \left(m\varphi\right) $ or $ -\!\sin \left(m\varphi\right) $, which we label by ``Re'' (real) or ``Im'' (imaginary) in the plots by convention, respectively. From those two options, we always select the nonvanishing modes for the plots.
    On the right-hand side, we visualize the corresponding deformational mode (always for positive value of the coefficient, negative coefficients invert this behavior). For $u^{\bot}$ in (a)--(c), we plot the deformed shape in blue in comparison to the undeformed shape in yellow. In the first two cases, we use side views, while the third row contains a top view as in Fig.~\ref{fig:star_configs}. For $u^{\theta}$ in (d,e), we color-code the value of the corresponding spherical harmonic in a side view of the sphere with the $x$-axis in the center. We illustrate the tangential polar displacements, given by $u^{\theta}$, which the mode corresponds to using black arrows. Black dots indicate negligible displacements at this position. Similarly, we plot $u^{\varphi}$ in (f), but now for corresponding tangential azimuthal displacements. 
    As we found only an imaginary mode here, we also plot the deformations corresponding to the imaginary part of this mode (superscript $Im$).
    In (c) and (f), we plot that mode of $l$ and $m$ set to the value $n_{arms}$ of the corresponding configuration. The visualizations on the right-hand side of those rows correspond to the case of $n_{arms}=4$.}
	\label{fig:star_results}
\end{figure}

Under magnetization, all these arrangements induce magnetostrictive deformations of the material. We here characterize and define the modes of deformation according to expansions into spherical harmonics. The most important modes, i.e.\ those of overall largest (absolute) magnitude, are plotted in Fig.~\ref{fig:star_results} for the geometries of different values of $n_{arms}$. Corresponding types of deformation are illustrated on the right-hand side of Fig.~\ref{fig:star_results}. On the one hand, for the modes associated with the component $u^{\bot}$, the overall shape of the material is affected and we plot what the new shape looks like if only this mode of positive sign were present. On the other hand, the modes associated with $u^{\theta}$ (and also $u^{\varphi}$) describe displacements tangential to the surface. Therefore, the overall shape of a sphere is not changed by them. Instead, possibly local, tangential displacements are included by them.
For modes related to the spherical harmonic $Y_{l,m}$ with $m \neq 0$, the resulting $\varphi$-dependence can manifest itself in two different ways: Either as $ \cos \left(m\varphi\right)  $ or as $ -\sin \left(m\varphi\right) $. In the former case, we write it (using the convention of the HEALPix package \cite{HEALPix}), as a real expansion coefficient, indicated by the letters ``Re'' in the plots. In the latter case, as an imaginary expansion coefficient, indicated by ``Im'' (with the negative sign by convention). Obviously, the resulting displacement field is real in both cases, which is a result of combining the terms with positive and negative $m$ -- we here only plot the expansion coefficient for positive $m$. In our plots, we only show the nonvanishing modes, i.e.\ for the plots with ``Re'', the corresponding imaginary part is approximately zero and vice versa. For further details, we refer to Ref.~\onlinecite{fischer2020}.
The corresponding types of deformation are illustrated on the right-hand side of Fig.~\ref{fig:star_results} by the black arrows. The lengths and directions of the arrows describe the displacements induced by only the corresponding mode, again for positive sign.
For negative sign of the corresponding mode, all the directions are simply reversed.
We illustrate the mode corresponding to either the real or to the imaginary part of the coefficient, depending on which one was found as the nonvanishing coefficient in the plots on the left-hand side.

As can be inferred from Fig.~\ref{fig:star_results}(a), magnetizing chain-like structures mainly leads to overall reduction in volume, i.e.\ $ u^{\bot}_{00} < 0$ (except for incompressible materials of $\nu = 0.5$). We note here that for generic visualization on the right-hand side, we there always refer to positive values of the mode. In the present case, the illustration on the right-hand side of Fig.~\ref{fig:star_results}(a) is for $ u^{\bot}_{00} > 0$. As we obtain negative values from our expansion, the material behavior is inverse to that, namely a reduction in volume instead of an increase in volume.
Additionally, we find $ u^{\bot}_{20} < 0$, see Fig.~\ref{fig:star_results}(b), which indicates a contraction along the magnetization direction (also referred to as oblate deformation). Again, on the right-hand side of Fig.~\ref{fig:star_results}(b), the visualization is for $ u^{\bot}_{20} > 0$, referring to an elongation along the magnetization direction. The observed behavior is expected from the attractive interactions along the magnetization direction of each chain-like aggregate. The mode $ u^{\theta}_{10} > 0$ in Fig.~\ref{fig:star_results}(d) indicates material displacements along the tangential direction on the surface of the sphere, within all planes containing the magnetization direction. Displacements are in opposite directions on the upper and lower hemisphere. This again contains the effect of a contraction along the magnetization direction, yet here tangentially to the surface. With lower magnitude, the values $ u^{\theta}_{30} < 0$ in Fig.~\ref{fig:star_results}(e) modulate this behavior, indicating that displacements towards the equatorial plane are more pronounced closer to it than near the poles.
Mostly, we observe an increasing trend of all these kinds of deformation with $n_{arms}$. This is in line with the increasing number of magnetizable inclusions. Simultaneously, the higher values of $n_{arms}$ decrease the distance between nearby chains, therefore pronouncing the repulsive interactions between neighboring inclusions from different chains.

Additionally, we plot two modes that are strongly related to the considered configurations. Namely, these are the modes $ u^{\bot}_{n_{arms},n_{arms}}$ and $ u^{\varphi}_{n_{arms},n_{arms}}$, where we set both values of $l$ and $m$ of the corresponding spherical harmonic $Y_{l,m}$ to $n_{arms}$. Therefore, these modes reflect the symmetry of the underlying spatial configuration. For example, in Fig.~\ref{fig:star_results}(c), we observe that for $n_{arms}=4$ the deformations characterized by $u^{\bot}_{44}$ show a four-fold symmetry as does the first configuration of $n_{arms}=4$ in the top view in Fig.~\ref{fig:star_configs}. In the vicinity of each chain located closer to the surface of the sphere, the contraction on the surface along the magnetization direction is more pronounced when compared to the positions that are further away from the chains. In general, this behavior is more obvious for low $n_{arms}$, for which the distance between different chains is larger and the deformation on the surface is therefore varying more significantly. In $ u^{\varphi}_{n_{arms},n_{arms}}$, we observe a similar behavior, see Fig.~\ref{fig:star_results}(f). This mode shows that the surface of the elastic sphere close to the ends of the chains is displaced towards the positions of the chains. For example, the arrows in Fig.~\ref{fig:star_results}(f) are all positioned at $\varphi = -\pi / 8 $ and indicate displacements towards $\varphi = 0$, where we find one of the chain-like aggregates corresponding to the end of one arm of our planar star-shaped arrangement.

The magnitude of most of these modes increase with decreasing Poisson ratio, that is, as the elastic matrix becomes more compressible and finally auxetic, see Fig.~\ref{fig:star_results}. This tendency is expected because a more compressible matrix less severely restricts the deformations. However, the trend is inverted for some mode of coefficient $ u^{\bot}_{20} < 0$, see Fig.~\ref{fig:star_results}(b) where the absolute magnitude decreases with decreasing Poisson ratio for $n_{star} \geq 8$, probably due to the more pronounced repulsive interactions as mentioned above. An inverted behavior is also observed in the modes of coefficients $ u^{\theta}_{10} > 0$ and $ u^{\theta}_{30} < 0$ for low values of $n_{star}$ in Fig.~\ref{fig:star_results}(d) and (e), respectively.

\subsubsection{Planar star-shaped arrangements without vertical shift between neighboring chains}
\label{subsubsec:star2}
	
\begin{figure}
	\includegraphics[width=\linewidth, trim={4.7cm 0.cm 5.cm 0},clip]{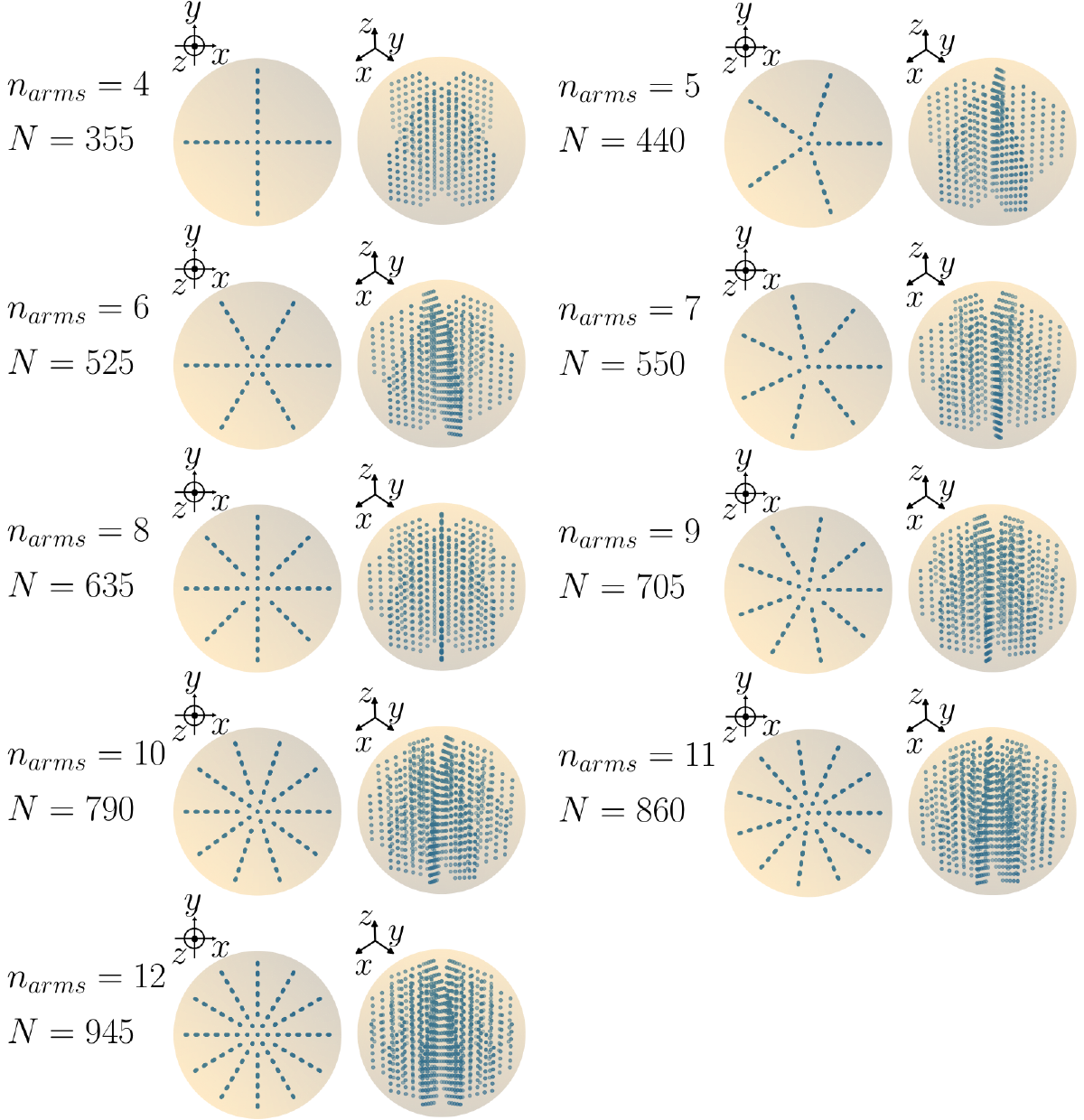}
	\caption{Top views (left of each pair of snapshots) and tilted top views (right of each pair) of the star-shaped arrangements of chains of particles similarly to the configurations displayed in Fig.~\ref{fig:star_configs}, but without mutual vertical shifts of neighboring chains along the magnetization direction $\bfhat{z}$. Again, $n_{arms}$ indicates the number of arms of each star and  $N$ sets the number of magnetizable inclusions. Appendix B contains further structural details on these configurations.}
	\label{fig:star_configs2}
\end{figure}

We also investigate planar star-shaped arrangements that were constructed in the same way as in the previous Sec.~\ref{subsubsec:star}, but without the mutual vertical shift of neighboring chains. Instead, we increase the distance between neighboring chains to $0.121 R$ so that neighboring inclusions still approximately maintain the minimal distance according to our constraints. For $n_{arms} \geq 7$, some chains close to the center again must be deleted, see also the top views of the configurations in Fig.~\ref{fig:star_configs2} and Appendix B.

\begin{figure}
	\includegraphics[width=\linewidth, trim={0.2cm 0.1cm 0.cm 0.cm},clip]{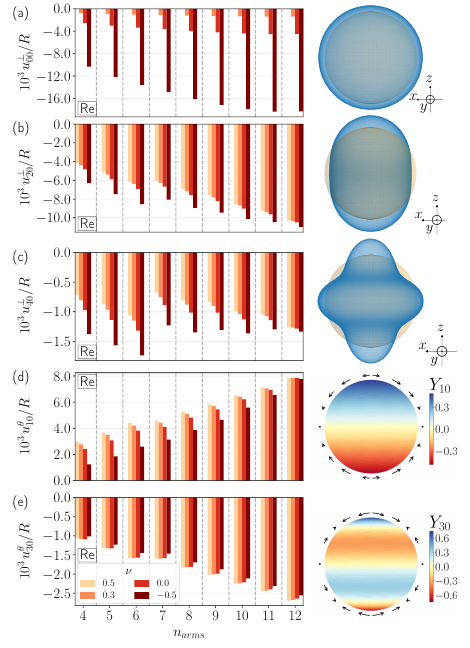}
	\caption{For the star-shaped chain configurations without mutual vertical shift of neighboring chains along the magnetization direction in Fig.~\ref{fig:star_configs2}, we evaluate the most important deformational modes and visualize them in the same way as in Fig.~\ref{fig:star_results}, varying the number of arms $n_{arms}$ of the star-shaped arrangements.}
	\label{fig:star_results2}
\end{figure}

In general, without mutual vertical shift the number of inclusions $N$ needs to be decreased (see Fig.~\ref{fig:star_configs} and Fig.~\ref{fig:star_configs2} for comparison). As may be expected, the magnetostrictive behavior is affected. We plot the most relevant modes of deformation in Fig.~\ref{fig:star_results2} in a similar manner as in the previous section in Fig.~\ref{fig:star_results}.

Fig.~\ref{fig:star_results2} already shows substantial differences compared to Fig.~\ref{fig:star_results}, while the configurations themselves in general are quite similar. We notice that the difference depends on the exact employed configuration, that is, the value of $n_{arms}$, and also on the Poisson ratio $\nu$.
In detail, we observe an almost uniform decrease of the dominant mode with coefficient $u^{\bot}_{00}$ that indicates changes in volume in Fig.~\ref{fig:star_results2}(a). Associated magnitudes are reduced to approximately $70$--$75~\%$ of the corresponding previous values in Fig.~\ref{fig:star_results}(a). 

In contrast to that, for the mode $u^{\bot}_{20}$ in Fig.~\ref{fig:star_results2}(b), the amplitudes are strongly increased (up to fourfold). 
Such behavior is in line with the more pronounced repulsive interactions between neighboring magnetized inclusions that now occupy identical planes. Here, $u^{\bot}_{20}<0$, so that the system contracts along the magnetization direction and extends in the planes of mutual magnetic repulsion. Besides, the auxetic systems ($\nu=-0.5$) always show the largest magnitude, while in the previous situation this was only the case for $n_{arms} \leq 7$.

In Fig.~\ref{fig:star_results}, the coefficient $u^{\bot}_{40}$ is not shown, but here the associated mode is a lot more significant and enters Fig.~\ref{fig:star_results2}(c). $u^{\bot}_{40}$ is always negative for the configurations without vertical shift, while for the previous configurations with vertical shift it features both signs. This mode of negative coefficient here represents displacements that are oriented inwards at the poles and in the equatorial plane while outwards otherwise. Its increase in absolute magnitude is about or more than twofold when excluding the vertical shift.

Next, the mode of coefficient $u^{\bot}_{n_{arms},n_{arms}}$ is strongly reduced in importance and at least twofold, partially fourfold, in magnitude when compared to the configurations with vertical shift. Therefore, it does not enter Fig.~\ref{fig:star_results2}. Also, the sign is sometimes opposite.

Concerning the mode of coefficient $u^{\theta}_{10}$ in Fig.~\ref{fig:star_results2}(d), it shows a slight decrease in magnitude in some cases, but mainly a strong increase up to $84~\%$, when compared to Fig.~\ref{fig:star_results}. When decreasing the Poisson ratio in Fig.~\ref{fig:star_results2}(d), this mode always decreases in magnitude.

Besides, for the mode of coefficient $u^{\theta}_{30}$ in Fig.~\ref{fig:star_results2}(e), we observe an increase that is more uniform, between approximately $26$--$64~\%$, when compared to the systems with vertical shift in Sec.~\ref{subsubsec:star}. In Fig.~\ref{fig:star_results2}(e), the auxetic systems always show the lowest absolute magnitude of this mode, nonmonotonic dependence on the Poisson ratio is partly found.

Lastly, the mode $u^{\varphi}_{n_{arms},n_{arms}}$ is generally substantially decreased, at least when compared to its high magnitudes in Fig.~\ref{fig:star_results}(f). For certain parameters, we do observe an increase, but only at a very low level. Therefore, this mode is less prominent when compared to the modes included in Fig.~\ref{fig:star_results2}.

\subsubsection{Circular arrangements}
\label{subsubsec:circle}
\begin{figure}
	\includegraphics[width=\linewidth, trim={0.2cm 2.5cm 0.5cm 2.3cm},clip]{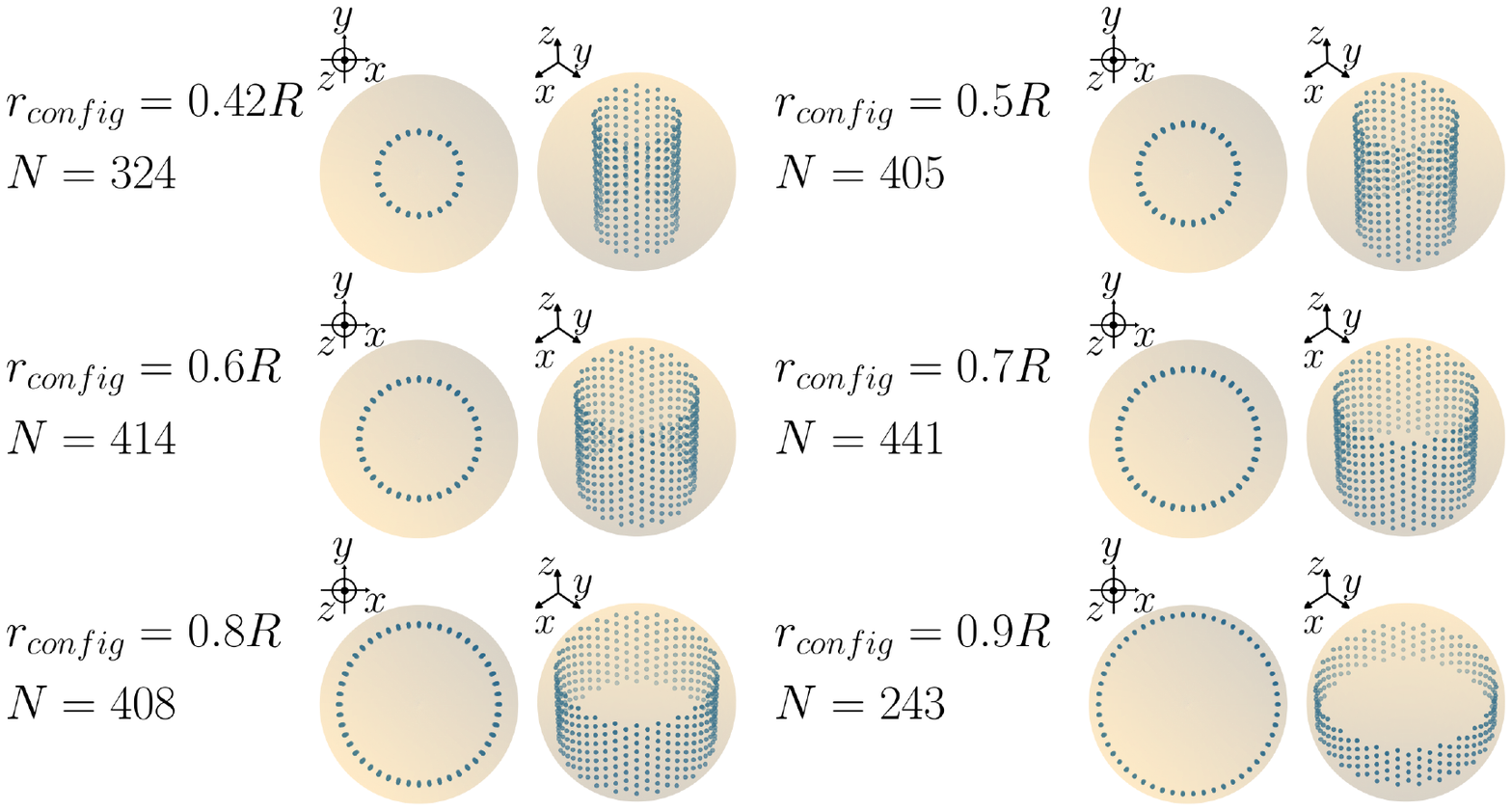}
	\caption{Top views (left of each pair of snapshots) and tilted top views (right of each pair) of the circular configurations of chain-like aggregates of magnetizable inclusions. The chains are oriented along the magnetization direction $\bfhat{z}$. $r_{config}$ sets the radius of the circle and $N$ sets the number of magnetizable inclusions.}
	\label{fig:circle_configs}
\end{figure}

Next, we address planar circular configurations of chains, with $r_{config}$ denoting the radius of each circular arrangement. Each circle in the central plane of the elastic sphere is centered at the center of the sphere. Chain-like aggregates are again built starting from this central plane as described in Sec.~\ref{subsec:chains}. As in Sec.~\ref{subsubsec:star}, a vertical shift between neighboring chains on the circle is then imposed as a further step. The procedure of filling the sphere is pursued in a way to in the end obtain that number of chains corresponding to the maximum number allowed under the constraints introduced before.
Resulting configurations are visualized in Fig.~\ref{fig:circle_configs}.

\begin{figure}
	\includegraphics[width=\linewidth, trim={0.2cm 0.1cm 0.cm 0.cm},clip]{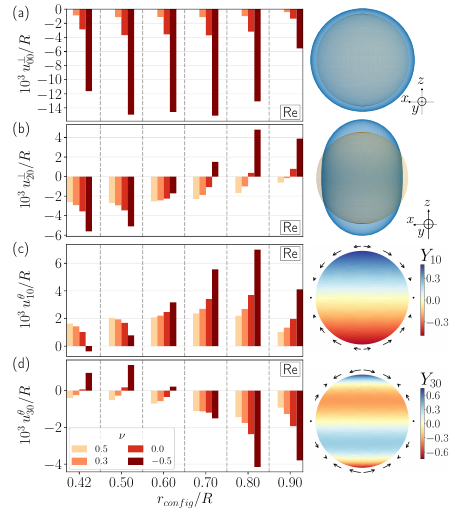}
	\caption{Magnitudes of the most important deformational modes as in Fig.~\ref{fig:star_results}, but for the circular configurations of Fig.~\ref{fig:circle_configs} with varying values of $r_{config}$.}
	\label{fig:circle_results}
\end{figure}

The induced overall deformations of the sphere when magnetizing these configurations are visualized in Fig.~\ref{fig:circle_results}, in analogy to Fig.~\ref{fig:star_results}. The results are similar to the results in Fig.~\ref{fig:star_results} and Fig.~\ref{fig:star_results2}.
Again, we find $ u^{\bot}_{00} < 0$ and $ u^{\theta}_{10} > 0$, except for auxetic materials ($\nu = -0.5$) and $r_{config}=0.42 R$ where we find $ u^{\theta}_{10} < 0$, see Fig.~\ref{fig:circle_results}(b) in the first column. Here, we note that the number of inclusions $N$ is not increasing monotonically with $r_{config}$, therefore, we cannot expect a monotonic increase in the magnitudes of the deformational modes. The behavior of the mode $ u^{\bot}_{20} $ is even changing sign with increasing $r_{config}$. For low $r_{config}$, the behavior is as expected for individual chain-like aggregates, and we find contractions along the magnetization direction, i.e.\ $ u^{\bot}_{20} < 0$. For larger $r_{config}$, the chains get shorter as the available space in the magnetization direction inside the elastic sphere decreases.
Thus, repulsive interactions perpendicular to the magnetization direction between inclusions part of different chains become more relevant. A similar change in sign can partly be observed in the mode $ u^{\theta}_{10} $ for $ \nu = -0.5 $ and in the mode $ u^{\theta}_{30} $ for $ \nu \leq 0 $.

\subsubsection{Concentric ring-like and tubular arrangements}
\label{subsubsec:concentric_rings_tube}

\begin{figure}
	\includegraphics[width=\linewidth, trim={0.2cm 7.5cm 0.5cm 7.2cm},clip]{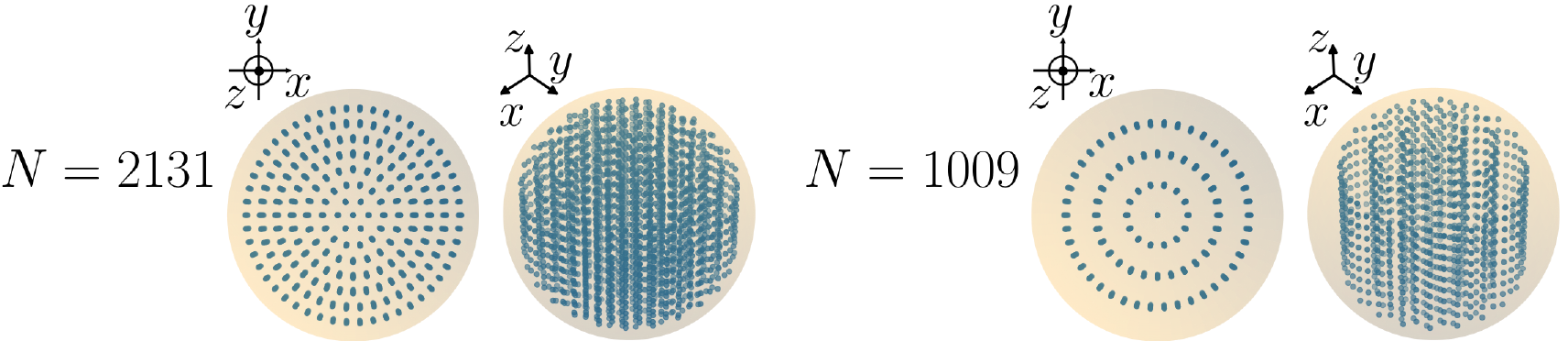}
	\caption{Top views (left of each pair of snapshots) and tilted top views (right of each pair) of the concentric ring-like configurations of chain-like aggregates of magnetizable inclusions. The chains are oriented along the magnetization direction $\bfhat{z}$.}
	\label{fig:concentric_ring_configs}
\end{figure}

Building on the results in Sec.~\ref{subsubsec:circle}, we now combine circles of different $r_{config}$ in one configuration. Two different variants are selected, one with a spacing of $0.121 R$ in the radial direction, i.e.\ as a spacing between different $r_{config}$, and one with twice that amount. Again, we place as a many chains as permitted by our restrictions on each circle, see Fig.~\ref{fig:concentric_ring_configs}.

\begin{figure}
    \centering
	\includegraphics[width=.7\linewidth, trim={0.3cm 0.2cm 0.6cm 0.1cm}, clip]{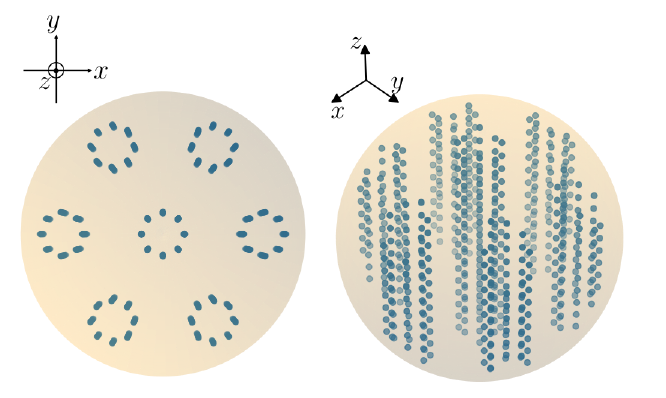}
	\caption{Top view (left) and tilted top view (right) of the tubular configuration of chain-like aggregates of $N = 600$ magnetizable inclusions. The chains are oriented along the magnetization direction $\bfhat{z}$.}
	\label{fig:tube_config}
\end{figure}

In tubular configurations, chains are still placed on circles. However, these circles are not concentric. Instead, $7$ circles are introduced in the equatorial plane, their centers arranged in a hexagonal manner. Each circle is of a radius of $0.15R$, with a center-to-center distance between the circles of $0.7R$ for nearest neighbors. From there, we grow from each inclusion in the equatorial plane a chain-like aggregate along the magnetization direction, which results in tubular configurations, see Fig.~\ref{fig:tube_config}. Related tubes were identified in x-ray tomographic investigations of real experimental samples of magnetic elastomers prepared under strong homogeneous external magnetic fields \cite{gunther2011x}.

\begin{figure}
	\includegraphics[width=\linewidth, trim={0.2cm 0.1cm 0.cm 0.cm},clip]{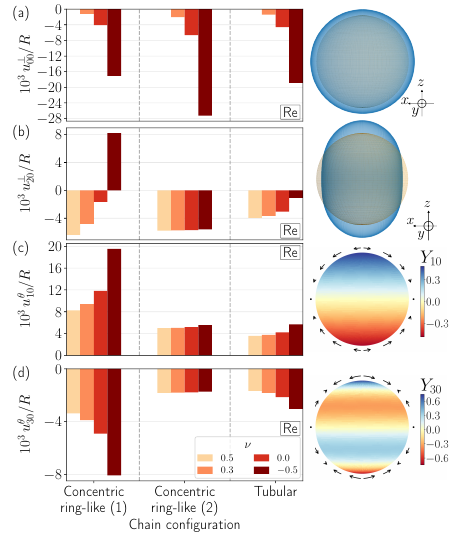}
	\caption{Magnitudes of the most important deformational modes as in Fig.~\ref{fig:star_results}, but for the concentric ring-like and tubular configurations of Figs.~\ref{fig:concentric_ring_configs} and \ref{fig:tube_config}, respectively.}
	\label{fig:concentric_ring_tube_results}
\end{figure}

Probably due to the attractions along each chain-like aggregate, we find results similar to those for the star-shaped arrangements depicted in Fig.~\ref{fig:star_results} and Fig.~\ref{fig:star_results2} for both concentric ring-like (first and second column of Fig.~\ref{fig:concentric_ring_tube_results}) and tubular (third column of Fig.~\ref{fig:concentric_ring_tube_results}) configurations. That is, volume changes are negative ($ u^{\bot}_{00} < 0$), contraction along the magnetization direction prevails ($ u^{\bot}_{20} < 0$), and qualitatively similar tangential response is observed ($ u^{\theta}_{10} > 0$, $ u^{\theta}_{30} < 0$). The notable exception to this list is  $ u^{\bot}_{20} < 0$ for the denser concentric ring-like arrangement. In this case, there are many chains close together,
which emphasizes the repulsive interactions perpendicular to the magnetization direction. Therefore, it appears conceivable that we observe elongation ($ u^{\bot}_{20} > 0$) along the magnetic field direction in this case at least for auxetic materials of $\nu = -0.5$, see the first column of Fig.~\ref{fig:concentric_ring_tube_results}(b).

\subsubsection{Polygonal arrangements}
\label{subsubsec:polygon}
\begin{figure}
	\includegraphics[width=\linewidth, trim={0.2cm 4.5cm 0.5cm 4.1cm},clip]{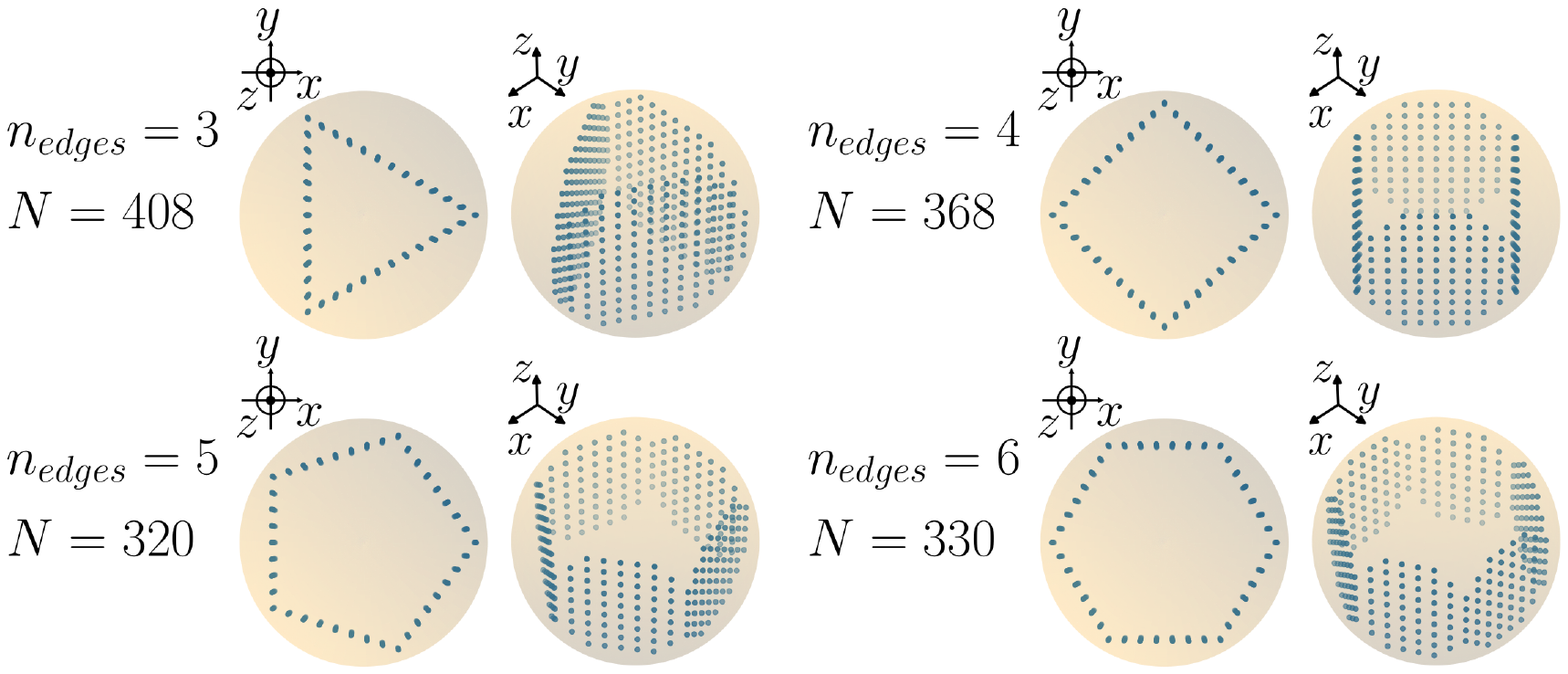}
	\caption{Top views (left of each pair of snapshots) and tilted top views (right of each pair) of the polygonal configurations of chain-like aggregates of magnetizable inclusions. The chains are oriented along the magnetization direction $\bfhat{z}$. $n_{edges}$ indicates the number of edges of each polygon.}
	\label{fig:polygon_configs}
\end{figure}

We also constructed chain-like aggregates from regular polygonal arrangements in the equatorial plane. The center of each polygon coincides with the center of the sphere. All vertices of each polygon therefore are located at identical distance from the center, which we here choose as $0.9R$. We show results for different numbers of polygonal edges, denoted by $n_{edges}$, which equals the number of vertices, see Fig.~\ref{fig:polygon_configs}.

\begin{figure}
	\includegraphics[width=\linewidth, trim={0.2cm 0.1cm 0.cm 0.cm},clip]{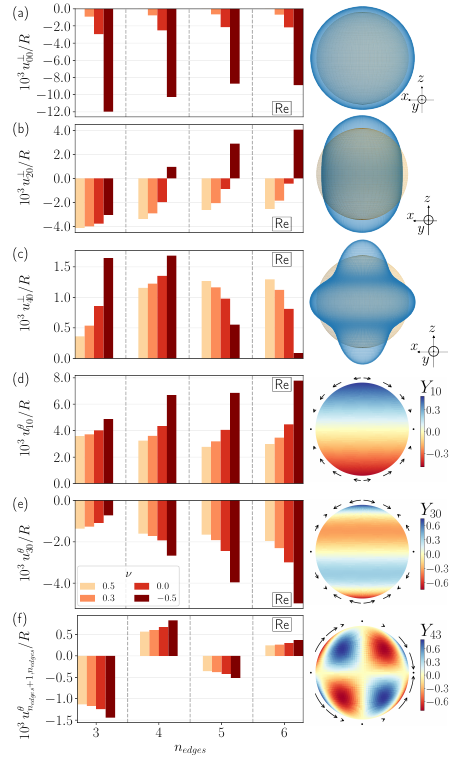}
	\caption{Magnitudes of the most important deformational modes as in Fig.~\ref{fig:star_results}, but for the polygonal configurations of Fig.~\ref{fig:polygon_configs} with varying values of $n_{edges}$. The visualization in (f) corresponds to the case $n_{edges}=3$.}
	\label{fig:polygon_results}
\end{figure}

For these arrangements, the induced modes of magnetostrictive deformation are presented in Fig.~\ref{fig:polygon_results}. Qualitatively, they are quite similar to the ones for circular arrangements in Fig.~\ref{fig:circle_results}.
As in the circular case for increasing $r_{config}$, increasing $n_{edges}$ on average moves the chains further away from the center of the elastic sphere and therefore shortens the length of the chains that fits into the sphere. This most likely leads to the decrease with increasing $n_{edges}$ in induced shrinkage of volume obvious from Fig.~\ref{fig:polygon_results}(a) and also to the decreasing contraction along the magnetization direction from Fig.~\ref{fig:polygon_results}(b). In the latter case, we even observe a change in sign for auxetic materials as in Fig.~\ref{fig:circle_results}(b). 

However, we here find $ u^{\theta}_{30} < 0$ for all evaluated parameters, which implies that tangential displacements towards the equatorial plane are decreasing in magnitude with increasing distance to the equatorial plane, qualitatively similar to the results in Fig.~\ref{fig:star_results}(e). Additionally, we also present the results for $ u^{\bot}_{40} $  in Fig.~\ref{fig:polygon_results}(c), where we infer that the displacements are more outwards at the poles and the equatorial plane and inwards otherwise. This modulation in the perpendicular component is analogous to the one that is represented by $ u^{\theta}_{30} < 0$ for the tangential component in Fig.~\ref{fig:polygon_results}(e).

We here also plot the mode $ u^{\theta}_{n_{edges}+1,n_{edges}}$ in Fig.~\ref{fig:polygon_results}(f).
The mirror-symmetry of the underlying configuration with respect to the equatorial plane implies an antisymmetric mode for $ u^{\theta}$. Therefore, the parameters $l$ and $m$ of any nonvanishing  mode corresponding to the spherical harmonic $Y_{l,m}$ have to be such that $l+m$ is odd. Consequently, the lowest-order mode for $ u^{\theta}$ that reflects the $n_{edges}$-fold symmetry in $m$ is the one that we plot in Fig.~\ref{fig:polygon_results}(f).

\subsubsection{Hexagonal arrangements}
\label{subsubsec:hexagon}

\begin{figure}
    \includegraphics[width=\linewidth, trim={0.1cm 3.7cm 0.5cm 3.8cm},clip]{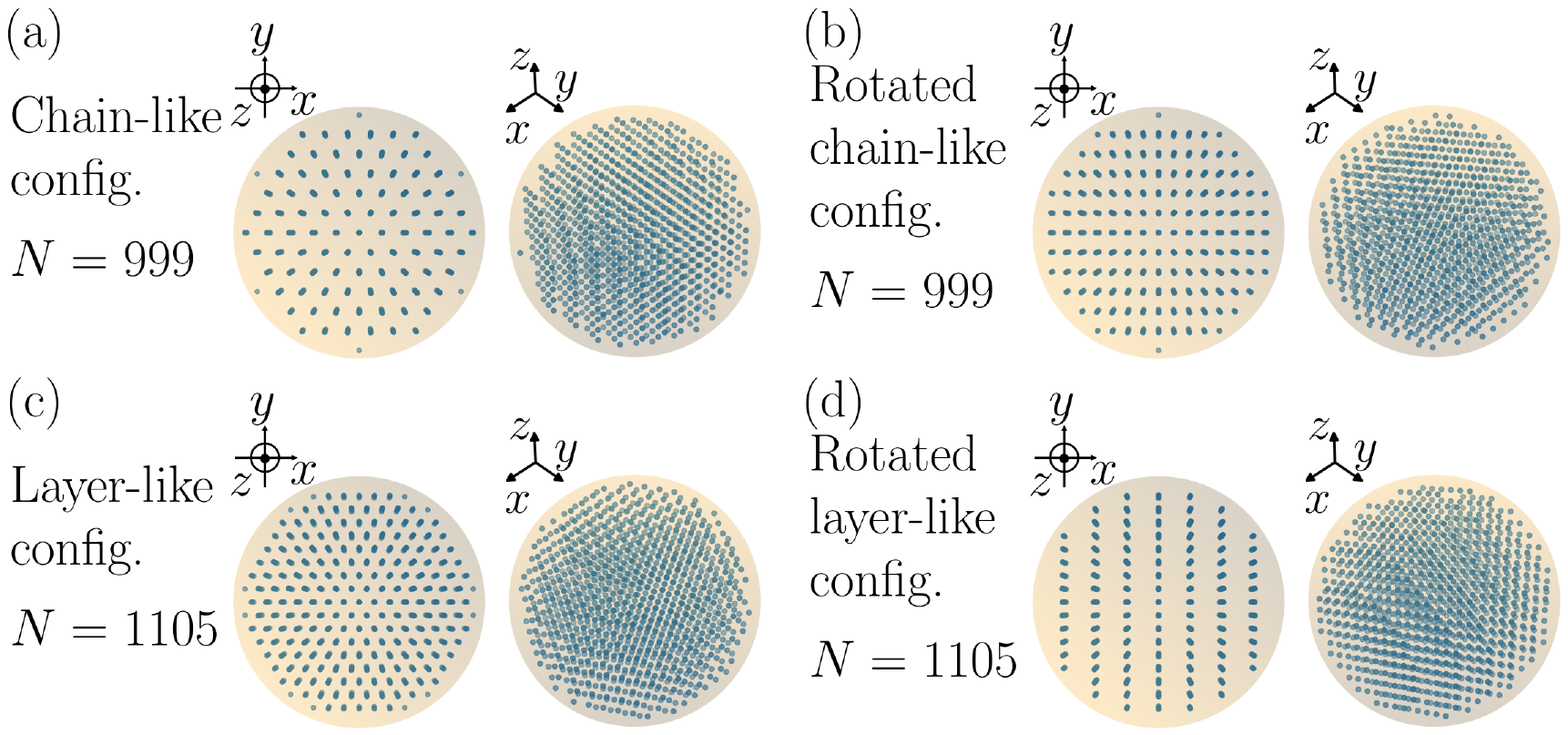}
	\caption{Top views (left of each pair of snapshots) and tilted top views (right of each pair) of the hexagonal configurations (configs.) of chain-like aggregates of magnetizable inclusions. The chain-like aggregates are oriented along the magnetization direction $\bfhat{z}$ in (a) and along $\bfhat{x}$ (pointing to the right) in (b). We also investigate stacks of hexagonally structured layers instead of hexagonally arranged chain-like aggregates (smaller inner-layer spacing between nearest-neighboring inclusions when compared to the inter-inclusion distance along the chain-like aggregates). In the latter case, we investigate layer structures of normal vector along (c) $\bfhat{z}$ and (d) $\bfhat{x}$.}
	\label{fig:hexagon_configs}
\end{figure}

Finally, we also consider simple hexagonal arrangements of chain-like aggregates oriented along the magnetization direction. Here, the distance of neighboring chains in the equatorial plane is set to $0.18 R$. Additionally, we consider the case in which the chain-like aggregates are oriented along the $x$-direction instead. To this end, we apply a global rotation matrix $\mathbf{R}$ to all position vectors of magnetizable inclusions, where
\begin{align} \mathbf{R}=
    \begin{pmatrix}
    0 & 0 & 1 \\
    0 & 1 & 0 \\
    -1 & 0 & 0
    \end{pmatrix}.
\end{align}
$\mathbf{R}$ maps the $x$-components to negative $z$-components and $z$-components to $x$-components, while leaving the $y$-components unaffected.
Moreover, we address configurations that feature an interinclusion distance of $0.25 R$ along the magnetization direction (compared to $0.121 R$ previously). Instead, we reduce the interinclusion spacing within the hexagonal plane to $0.121 R$. Therefore, we construct a configuration of hexagonally structured layers stacked on top of each other (normal vector $\bfhat{z}$). For these hexagonal layer structures, we also investigate the rotated configurations (normal vector $\bfhat{x}$) after applying the same rotation matrix $\mathbf{R}$. These four configurations are visualized in Fig.~\ref{fig:hexagon_configs}. In all cases, $\bfhat{z}$ still represents the magnetization direction.

\begin{figure}
	\includegraphics[width=\linewidth, trim={0.2cm 0.1cm 0.cm 0.cm},clip]{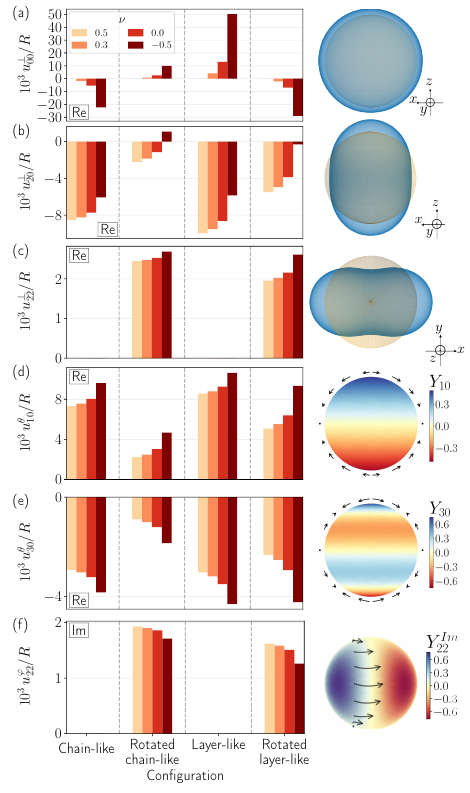}
	\caption{Magnitudes of the most important deformational modes as in Fig.~\ref{fig:star_results}, but for the hexagonally arranged chain-like aggregates and hexagonally structured layer-like configurations of Fig.~\ref{fig:hexagon_configs} with two different orientations for each case with the major axes along and rotated perpendicular to the magnetization direction.}
	\label{fig:hexagon_results}
\end{figure}

For the hexagonally arranged chain-like aggregates with chains along the magnetization direction (see the first column of Fig.~\ref{fig:hexagon_results}), we find a qualitatively similar magnetostrictive response as for the star-shaped arrangements of chains. A reduction in volume ($ u^{\bot}_{00} < 0 $) in connection with a contraction along the magnetization direction ($ u^{\bot}_{20} < 0 $) are observed, together with similar tangential deformations as for the aforementioned structures ($ u^{\theta}_{10} > 0 $, $ u^{\theta}_{30} < 0 $). Evidently, a rotation of the whole configuration with respect to the magnetization direction (see the second column of Fig.~\ref{fig:hexagon_results}) changes the behavior even qualitatively. The overall volume is now growing under magnetization ($ u^{\bot}_{00} > 0 $), as we might expect from the now predominately repulsive interactions within the perpendicularly magnetized chain-like aggregates.
We still find contraction along the magnetization direction ($ u^{\bot}_{20} < 0 $), or put differently, a sphere that expands upon magnetization across the equatorial plane, at least for $\nu \geq 0$. This also leads to the same qualitative behavior of $ u^{\theta}_{10} $ and $ u^{\theta}_{30} $. Moreover, due to the chains being oriented along the $x$-direction, we find additional pronounced modes in this case, namely $ u^{\bot}_{22} > 0 $, see Fig.~\ref{fig:hexagon_results}(c). This reflects the mirror-symmetry of the configuration with respect to the $yz$-plane. As we infer, the sphere expands in a more pronounced way along the $x$-axis. This is the axis along which the chains are oriented and, thus, the magnetic repulsion is strongest. Moreover, we observe $ u^{\varphi}_{22} > 0 $, see Fig.~\ref{fig:hexagon_results}(f), which indicates azimuthal displacements towards the $x$-axis (middle yellow part in the visualization of the mode).

Likewise, the hexagonal layer-like structures feature repulsive internal interactions as the layer normals are oriented along the magnetization direction (third column of Fig.~\ref{fig:hexagon_results}). Therefore, the qualitative response of the modes in many aspects is similar to the rotated chain-like configurations. As one exception, we find contraction ($ u^{\bot}_{20} < 0 $) in all cases, also in the auxetic one, see Fig.~\ref{fig:hexagon_results}(b). 

For the rotated configurations of hexagonal layer-like configurations, see the fourth column of Fig.~\ref{fig:hexagon_results}, upon magnetization we obtain repulsive interactions between nearby inclusions along the $y$-direction and attractive interactions along the $z$-direction. Both of these interactions contribute to a contraction along the magnetization direction ($ u^{\bot}_{20} < 0$), see Fig.~\ref{fig:hexagon_results}(b), as discussed previously. Again, we also observe $ u^{\theta}_{10} > 0 $ and $ u^{\theta}_{30} < 0$.  $ u^{\bot}_{00} $ here is found negative in Fig.~\ref{fig:hexagon_results}(a), implying an overall shrinkage of the elastic sphere.
We observe the same qualitative response in the modes $  u^{\bot}_{22} > 0 $ and $ u^{\varphi}_{22} > 0 $ as for the rotated chain-like aggregates, see Fig.~\ref{fig:hexagon_results}(c) and (f).

To summarize the above results, for chain-like aggregates arranged in different configurations but oriented along the magnetization direction, we mostly find an overall reduction in volume ($ u^{\bot}_{00} < 0 $) and contraction along the magnetization direction ($ u^{\bot}_{20} < 0 $). We usually observe tangential displacements towards the equatorial plane ($ u^{\theta}_{10} > 0 $). A majority of these modes increase in magnitude with decreasing Poisson ratio. Next, we concentrate on further configurations of magnetizable inclusions that do not consist of chain-like aggregates.

\subsection{Spherical arrangements}
\label{subsec:sphere}

\begin{figure}
	\includegraphics[width=\linewidth, trim={0.2cm 2.5cm 0.5cm 2.3cm},clip]{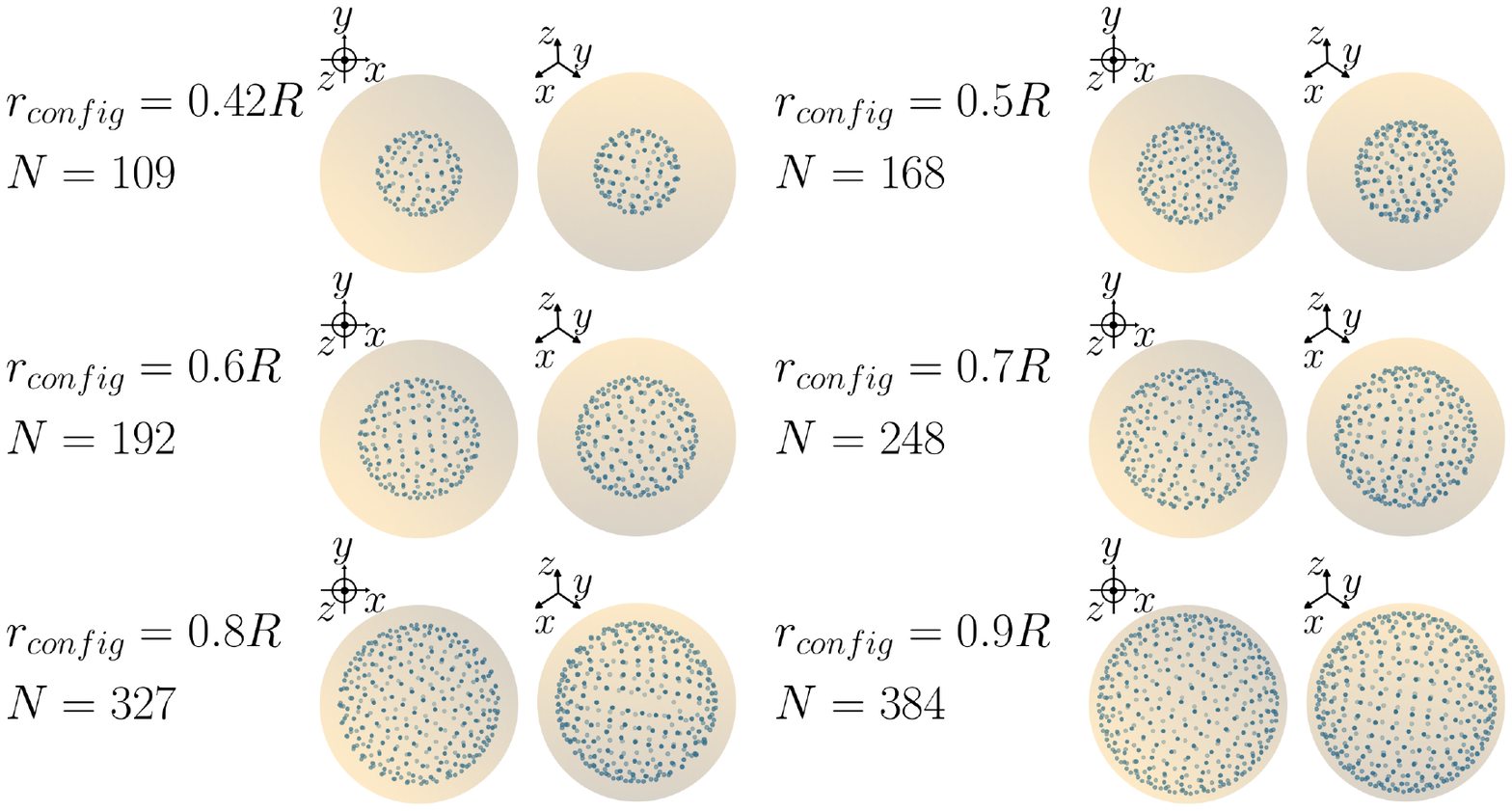}
	\caption{Top views (left of each pair of snapshots) and tilted top views (right of each pair) of the evenly distributed spherical arrangements of magnetizable inclusions, with $r_{config}$ as the radius of the sphere on which the inclusions are located.}
	\label{fig:sphere_configs}
\end{figure}

We continue by placing inclusions on the surface of spheres of different radii $r_{config} < R$ with the same parameters values for $r_{config}$ as in Sec.~\ref{subsubsec:circle}.
The inclusions are distributed approximately evenly on the sphere of radius $r_{config}$, which we ensure by use of the distributions provided by the HEALPix package \cite{HEALPix}, see Fig.~\ref{fig:sphere_configs} for the resulting configurations. We further ensure that the points of evaluating the displacement field on the surface of the overall elastic sphere of radius $R$ are not located on the same radial axis as the magnetizable inclusions underneath the surface of the sphere at distance $r_{config}$ from the center.

\begin{figure}
	\includegraphics[width=\linewidth, trim={0.2cm 0.1cm 0.cm 0.cm},clip]{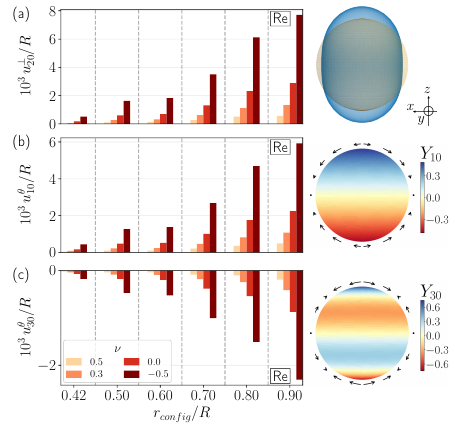}
	\caption{Magnitudes of the most important deformational modes as in Fig.~\ref{fig:star_results}, but for the spherical configurations of Fig.~\ref{fig:sphere_configs}.}
	\label{fig:sphere_results}
\end{figure}

As a result, we infer that the magnetostrictively induced changes in overall volume, characterized by the mode $ u^{\bot}_{00}$, are not dominant when compared to the other modes. Their maximal magnitudes are below $10^{-3}$ with an average magnitude below $10^{-4}$ and thus about one magnitude below those of the other modes. As the configurations are approximately isotropic, we find both attractive and repulsive magnetic interactions.
We infer from Fig.~\ref{fig:sphere_results} an elongation along the magnetization direction ($ u^{\bot}_{20} > 0 $, prolate deformation), in contrast to the contraction that was usually observed for the chain-like arrangements in Sec.~\ref{subsec:chains}. This result can be illustratively explained from the configuration of the inclusions. As we get closer to the equatorial plane, the nearest neighbors are increasingly located relative to each other along the magnetization direction. Consequently, they interact in a rather attractive manner. The opposite is true closer to the poles where repulsion dominates. This supports $ u^{\bot}_{20} > 0 $ and $ u^{\theta}_{10} > 0 $, see Fig.~\ref{fig:sphere_results}(b).
We also observe $ u^{\theta}_{30} < 0 $ in Fig.~\ref{fig:sphere_results}(c), qualitatively similar to Figs.~\ref{fig:star_results}(e) and \ref{fig:polygon_results}(e).
The magnitudes of deformation increase with increasing $r_{config}$, which again is in line with the increasing number of magnetizable inclusions $N$. The closer the magnetizable inclusions are located with respect to the elastic surface, that is, the larger $r_{config}$, the stronger the induced deformations of the surface.

\subsection{3D star-shaped arrangements}
\label{subsec:3d-star}
\begin{figure}
	\includegraphics[width=\linewidth, trim={0.2cm 7.6cm 0.5cm 7.3cm},clip]{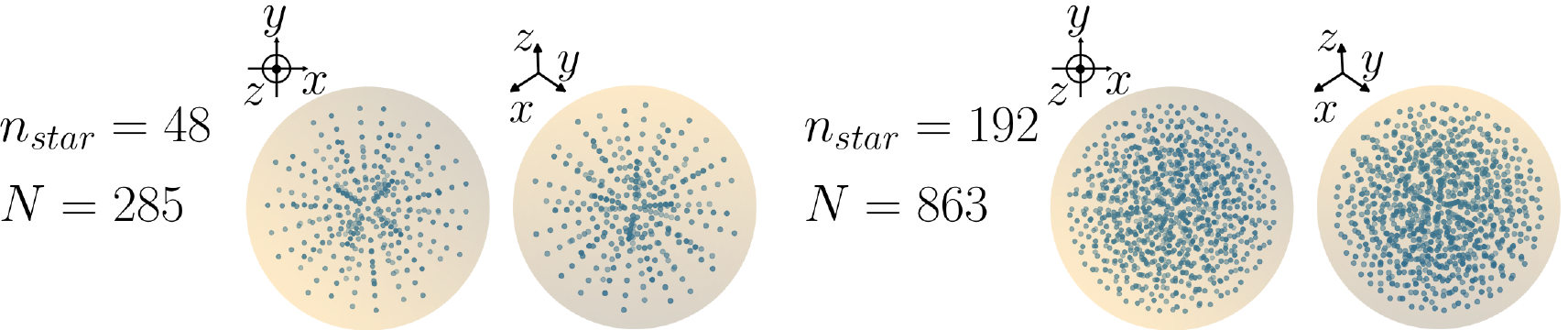}
	\caption{Top views (left of each pair of snapshots) and tilted top views (right of each pair) of the 3D star-shaped arrangements, where we consider stars with arms/chains oriented along the radial direction. They are directed towards a set of $n_{star}$  points that are approximately evenly distributed underneath the spherical surface.}
	\label{fig:3d-star_configs}
\end{figure}

In this section, we again distribute magnetizable inclusions evenly on a spherical surface as a starting point. However, we here set a lower number ($n_{star}=48$ or $n_{star}=192$ compared to the number of inclusions in Sec.~\ref{subsec:sphere}).
Then, we build chain-like aggregates from the center of the elastic sphere towards these locations up to the maximal number of inclusions that is permitted by our constraints. Thus, when compared to Sec.~\ref{subsubsec:star}, we now consider star-shaped configurations of chain-like aggregates arranged along the radial direction instead of the magnetization direction, see Fig.~\ref{fig:3d-star_configs}. The nearest-neighbor distance along the radially arranged chain-like aggregates is set to $0.121 R$ as before.

\begin{figure}
	\includegraphics[width=\linewidth, trim={0.2cm 0.1cm 0.cm 0.cm},clip]{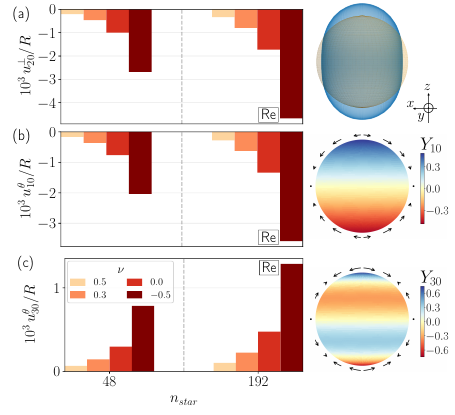}
	\caption{Magnitudes of the most important deformational modes as in Fig.~\ref{fig:star_results}, but for the 3D star-shaped arrangements of Fig.~\ref{fig:3d-star_configs}.}
	\label{fig:3d-star_results}
\end{figure}

When compared to the spherical configurations in the previous Sec.~\ref{subsec:sphere}, we again find the same most relevant modes. Likewise, the 3D star-shaped arrangements do not lead to relevant changes in volume. In contrast to the results in Fig.~\ref{fig:sphere_results}, here all further displayed modes are of opposite sign. We observe in Fig.~\ref{fig:3d-star_results}(a) a contraction along the magnetization direction ($ u^{\bot}_{20} < 0 $) as well as in Fig.~\ref{fig:3d-star_results}(b) and (c) $ u^{\theta}_{10} < 0 $ and $ u^{\theta}_{30} > 0 $.
In Fig.~\ref{fig:3d-star_results}, we again observe that an increase in $n_{star}$, coinciding with an increase in the number of magnetizable inclusions, increases the magnitudes of deformation.

\subsection{Single- and double-stranded helical arrangements}
\label{subsec:helical}

\begin{figure}
	\includegraphics[width=\linewidth, trim={0cm 4.4cm 0.5cm 4.3cm},clip]{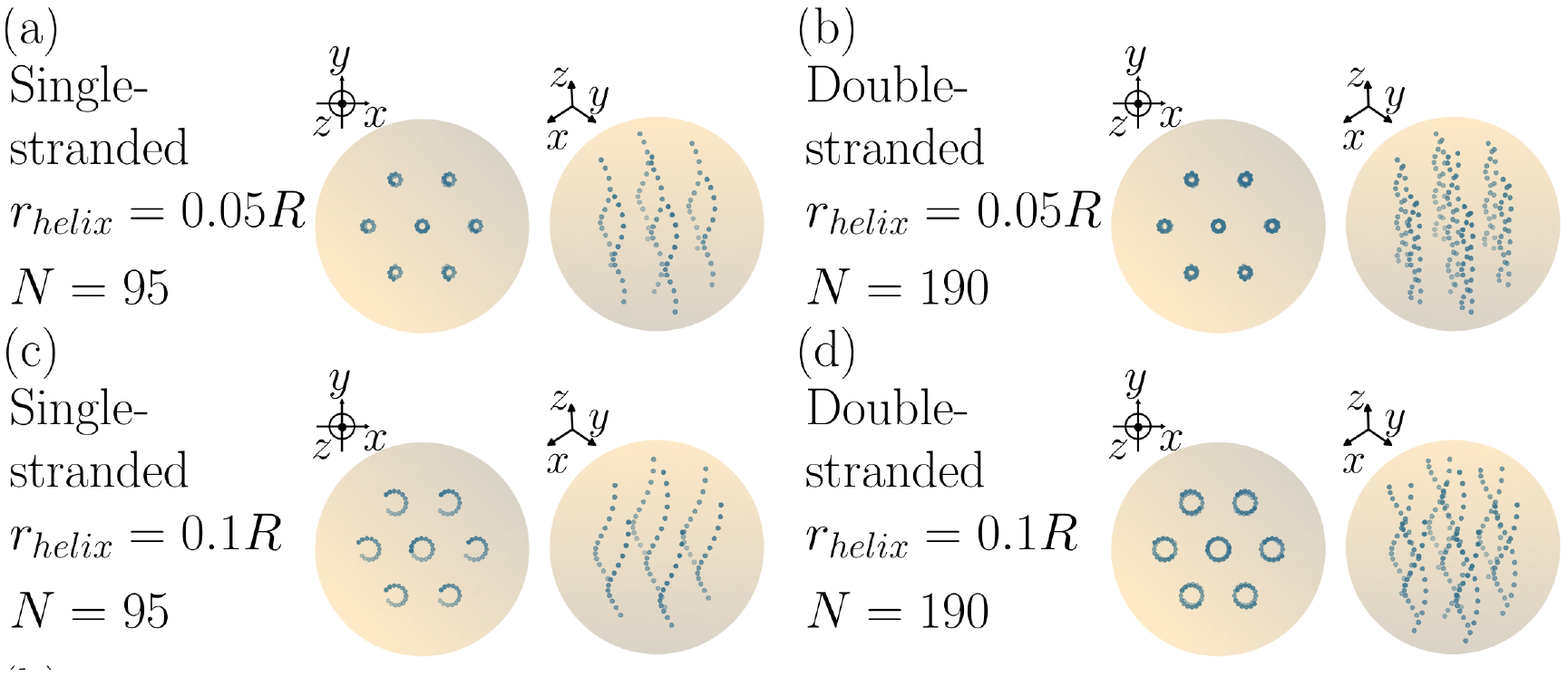}
	\caption{Top views (left of each pair of snapshots) and tilted top views (right of each pair) of the single- and double-stranded helical configurations. (a) and (c) correspond to configurations already explored in Ref.~\onlinecite{fischer2020}. We here choose those configurations featuring the largest twist-type deformations. The parameter $r_{helix}$ describes the radius of each helix. For the double-stranded helical configurations in (b) and (d), we add at the height of each inclusion another inclusion on the opposite side of the helical center axis.}
	\label{fig:helical_configs}
\end{figure}

Finally, we return to helical arrangements \cite{fischer2020}. Here, we extend the previous single-stranded \cite{fischer2020} to double-stranded helical configurations. One motivation in this context is to generate twist-type deformations, characterized by the azimuthal deformational mode $u^{\varphi}_{10} $. From above, these configurations appear similar to the tubular configurations of Sec.~\ref{subsubsec:concentric_rings_tube}, see Figs.~\ref{fig:tube_config} and \ref{fig:helical_configs}. 

We consider helical arrangements with center axes oriented along the magnetization direction. They are arranged on a hexagonal grid in the plane normal to the magnetization direction. The lattice constant of the hexagonal arrangement is chosen as $0.5R$. 
From each center point of the hexagonal lattice, we place one inclusion at a distance vector $\mathbf{r}_{helix}^{\bot}(z)$ in the plane perpendicular from the center axis of
\begin{align} \label{eq_helix_vec}
	\mathbf{r}_{helix}^{\bot}(z) &= r_{helix} \begin{pmatrix}
	\cos\frac{\gamma z}{ d_{layer}}\\[5pt]
	\sin\frac{\gamma z}{ d_{layer}}
	\end{pmatrix},
\end{align}
depending on the $z$-coordinate which we increase or decrease in discrete steps of $0.11 R$ from the central plane to fill the elastic sphere. We here discuss two choices for $r_{helix}$, namely $0.05 R$ and $0.1R$, the same values as in Ref.~\onlinecite{fischer2020}. For both cases, we choose the value of $\gamma$ that leads to the maximal overall twist-type response of the materials for single-stranded helices, namely $\gamma \approx 0.24 \pi$ and $\gamma \approx 0.13 \pi$ for $r_{helix}=0.05R$ and $r_{helix}=0.1R$, respectively. 

To expand on Ref.~\onlinecite{fischer2020}, we now also discuss the case of double-stranded helical configurations, inspired by the structure of DNA.
These configurations are generated by inserting additional inclusions at each step in the $\mathbf{\hat{z}}$-direction at $-\mathbf{r}_{helix}(z)$.
We note that for $r_{helix}=0.05R$ the distance between the two strands is $0.1 R < 0.12 R$ and thus slightly below our generally imposed minimal distance between any two inclusions of radius $0.02R$.
Besides, the double-stranded helical configurations contain twice the amount of magnetizable inclusions, see Fig.~\ref{fig:helical_configs} for the resulting configurations.

\begin{figure}
	\includegraphics[width=\linewidth, trim={0.2cm 0.1cm 0.cm 0.cm},clip]{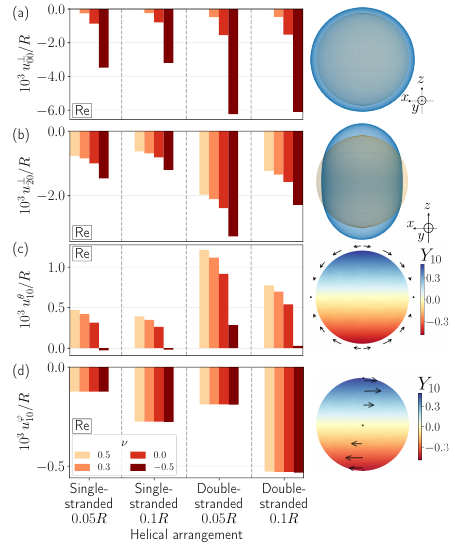}
	\caption{Magnitudes of the most important deformational modes as in Fig.~\ref{fig:star_results}, but for the single- and double-stranded helical arrangements depicted in Fig.~\ref{fig:helical_configs}.}
	\label{fig:helical_results}
\end{figure}

As we observe, the deformational response for the helical configurations (see the first two columns of Fig.~\ref{fig:helical_results}) shows a pattern familiar from the chain-like arrangements, namely a magnetostrictive reduction in volume ($ u^{\bot}_{00} < 0 $, except for $\nu=0.5$), contraction along the magnetization direction ($ u^{\bot}_{20} < 0 $) and mainly tangential displacements towards the equatorial plane ($ u^{\theta}_{10} > 0 $), see Fig.~\ref{fig:helical_results}(a), (b), and (c), respectively. In the case of auxetic materials of $\nu=-0.5$, we find $ u^{\theta}_{10} \approx 0 $. Overall, these characteristics appear conceivable because from a coarser point of view, the configurations to some extent resemble chain-like arrangements. Yet, in addition, all these configurations generate an overall twist-type deformational response as quantified by $u^{\varphi}_{10} $ in Fig.~\ref{fig:helical_results}(d). The magnitude of this mode is almost independent of the Poisson ratio, as also discussed in Ref.~\onlinecite{fischer2020}.

Concerning the newly addressed double-stranded helical configurations, in general, the magnitudes of all deformational modes depicted in  Fig.~\ref{fig:helical_results} are increased when compared to the single-stranded ones without changing the qualitative behavior. A slight exception is given by $ u^{\theta}_{10} > 0 $ for $\nu = -0.5$, see Fig.~\ref{fig:helical_results}(c). In particular, concerning the resulting twist-type deformation, $ u^{\varphi}_{10} $ is increased in magnitude by approximately $52.7\%$ and $92.2\%$ for $r_{helix}=0.05R$ and $r_{helix}=0.1R$, respectively, see Fig.~\ref{fig:helical_results}(d). Consequently, the double-stranded helical configurations are preferred relatively to the single-stranded ones, if a twist-type deformation of stronger extent is desired and the number of magnetizable inclusions is not restricted.

\section{Conclusions}
\label{sec:conclusions}

In this work, we demonstrate that specific modes of overall macroscopic deformation of magnetoelastic composite materials can be directly calculated from the specific discrete microscopic force pattern arising upon magnetization. This force pattern results from the spatial configuration and arrangement of the particulate inclusions and their mutual magnetic interactions. Our approach is based on analytical theory. It allows to effectively calculate the resulting elastic surface displacements and thus overall deformations by numerical evaluations. 

Thus, we show that a broad range of different spectra of modes of deformation can be realized by implementing different types of microscopic spatial arrangements of the particulate inclusions. For example, star-shaped arrangements of chain-like aggregates of a specific number of arms of the stars imply accordingly pronounced modes of overall deformation. Thus, if specific patterns of actuation are needed for specific purposes, for instance, filling of specifically shaped cavities by actuation of valves \cite{bose2012soft}, magnetoelastic composite materials are candidates to achieve such realizations. 

Experimentally, various ways of generating the different discrete particle arrangements need to be developed. For more macroscopic systems or macroscopic objects of demonstration, placement by molds or by hand is possible \cite{puljiz2016forces,puljiz2018reversible,zhang2008analysis,chen2013numerical}. When placing the inclusions by hand, the reactive polymeric suspension would be added layer by layer between successive events of placing inclusions. The newly forming elastic matrix develops by chemical reaction and connects to the already existing elastic part of the sample \cite{puljiz2016forces, puljiz2018reversible}. Pouring another layer into an overall spherical cavity to generate spherical samples may be difficult for the upper hemisphere when simultaneously adding inclusions by hand. Yet, the two hemispheres could be generated separately and then in the end be linked to each other by a final reaction.

3D-printing is a promising tool to be further developed in the future \cite{qi20203d,zhang20214d,bayaniahangar20213,kim2018printing, dohmen2020field,bastola2020dot,bayaniahangar20213,bastola20173d} to generate samples of controlled positioning of magnetizable inclusions. Here, one way to fabricate spherical systems would be to choose three types of ``inks''. The first one prints the elastic material that forms the deformable elastic sphere, representing the carrier matrix. From the second one, the magnetizable inclusions are positioned. Finally, the third ink could form a block of surrounding elastic gel material that embeds the elastic sphere. It should be much softer than the printed elastic sphere so that it hardly hinders the magnetostrictive deformation of the embedded spherical sample upon magnetization. It could also be generated from a material that can be selectively dissolved in a solvent after printing the whole arrangement \cite{komp2005versatile}, leaving the spherical sample at the end.

Additional methods of placing the inclusions in a requested manner are conceivable.
Prior to polymerization, the particle arrangement can also be controlled by external magnetic \cite{martin2006magnetostriction} or acoustic \cite{peerfischer} fields. Once the positioning is complete as desired, polymerizing the carrier matrix fixes their locations. Generally, polymerization could be performed stepwise using processes of photopolymerization to fabricate the matrix material \cite{kim2011programming}.

If the spherical samples are contained in a very soft surrounding transparent elastic box, see above, direct measurements are conceivable on such combined systems. Otherwise, free-standing elastic spheres can be maintained in a density-matched surrounding transparent fluid to then study magnetostrictive effects \cite{gollwitzer2008measuring}. A strong homogeneous external magnetic field should then be applied for experimental detection. The resulting magnetostrictive deformations can be recorded by standard optical tools \cite{gollwitzer2008measuring}.

Overall, we establish theoretical means that allow to effectively calculate the magnetomechanical response as a function of the microscopic particulate structure. We provide a palette of different such microscopic particle arrangements together with the magnetically inducible overall deformation. From this palette, specific realizations can be selected according to a particular need. In this way, we promote the application of magnetic gels and elastomers as tailored soft actuators adjusted to given requirements that are addressed and excited reversibly in a contactless fashion from outside by external magnetic fields. 

\section*{Data availability}
The generated data underlying the presented figures are published on the repository Zenodo and can be found at doi.org/10.5281/zenodo.10035624.

\begin{acknowledgments}

A.M.M. thanks the German Research Foundation (Deutsche Forschungsgemeinschaft, DFG) for support through the Heisenberg Grant no.\ ME 3571/4-2.
Some of the results in this paper have been derived using the HEALPix package \cite{HEALPix}.

\end{acknowledgments}

\section*{Appendix A: Details on the planar star-shaped configurations considered in Sec.~\ref{subsubsec:star}}
\label{sec:appendix1}
Here, we list explicitly the chains that we had to exclude from the configuration in Sec.~\ref{subsubsec:star} to satisfy our constraints in detail. 
For $n_{arms}=4$ and $n_{arms}=5$, no chains have been deleted.
If we number the arms of the star-shaped arrangements from $1$ to $n_{arms}$, with $1$ being the arm pointing to the right (along the $x$-axis) and then numbering them counter clockwise (mathematically positive sense of rotation), for $n_{arms}=6$, the arms $2$, $4$, and $6$ are missing the innermost chains. 
For $n_{arms}=7$, the innermost chains were deleted from the arms $2$, $4$, $6$, and $7$. 
Similarly, for $n_{arms}=8$, the innermost chains were deleted from the arms $2$, $4$, $6$, and $8$.
Likewise, for $n_{arms}=9$, the arms $2$, $4$, $6$, $8$, and $9$ are missing the innermost chains.
Again, for $n_{arms}=10$, we deleted from the arms $2$, $4$, $6$, $8$, and $10$ the innermost chains. 
For $n_{arms}=11$, the arms $2$, $3$, $5$, $6$, $8$, $9$, $10$, and $11$ lack the innermost chains and the arms $2$, $4$, $6$, $8$, $10$, and $11$ lack the second-innermost chains.
Finally, for $n_{arms}=12$, the arms $2$, $3$, $5$, $6$, $8$, $9$, $11$, and $12$ are missing the innermost chains and the arms $2$, $4$, $6$, $8$, $10$, and $12$ are missing the second-innermost chains.

\section*{Appendix B: Details on the star-shaped configurations without vertical shift considered in Sec.~\ref{subsubsec:star2}}
\label{sec:appendix2}
For the star-shaped configurations without vertical shift considered in Sec.~\ref{subsubsec:star2}, we likewise summarize which chains were deleted. We use the same notation as in Appendix A.
For $n_{arms}=4$ to $n_{arms}=6$, no chains are missing.
Next, for $n_{arms}=7$ to $n_{arms}=10$, the same chains are missing as in the case with vertical shift, see Appendix A.
For $n_{arms}=11$, the arms $2$, $4$, $6$, $8$, $10$, and $11$ are missing the innermost chains.
Finally, for $n_{arms}=12$, from the arms $2$, $4$, $6$, $8$, $10$, and $12$ we removed the innermost chains.

\bibliography{literature}

\end{document}